\documentclass[preprint,12pt]{aastex}
\usepackage{epsfig}
\usepackage{emulateapj5}
\usepackage{apjfonts}
\usepackage{graphicx}
\usepackage{rotating}
\usepackage{threeparttable,multirow}

\newcommand{\chandra}{\textit{Chandra}}

\newcommand{\flux}{\thinspace\hbox{$\hbox{ergs}\thinspace\hbox{cm}^{-2}\thinspace\hbox{s}^{-1}$}}
\newcommand{\hst}{\textit{HST}}

\makeatletter
\newenvironment{inlinetable}{%
\def\@captype{table}%
\noindent\begin{minipage}{0.999\linewidth}\begin{center}\footnotesize}
{\end{center}\end{minipage}\smallskip}

\newenvironment{inlinefigure}{%
\def\@captype{figure}%
\noindent\begin{minipage}{0.999\linewidth}\begin{center}}
{\end{center}\end{minipage}\smallskip}

\begin{document}

\def\spose#1{\hbox to 0pt{#1\hss}}
\def\laeq{\mathrel{\spose{\lower 3pt\hbox{$\mathchar"218$}}
     \raise 2.0pt\hbox{$\mathchar"13C$}}}
\def\gaeq{\mathrel{\spose{\lower 3pt\hbox{$\mathchar"218$}}
     \raise 2.0pt\hbox{$\mathchar"13E$}}}

\slugcomment{September 11, 2009, Accepted for Publication in the Astrophysical Journal}

\title{X-ray Sources and Their Optical Counterparts in the Galactic Globular Cluster M12 (NGC 6218)} 

\author{Ting-Ni~Lu\altaffilmark{1}, Albert~K.~H.~Kong\altaffilmark{1, 8},  Cees~Bassa\altaffilmark{2, 3}, Frank~Verbunt\altaffilmark{4}, 
Walter~H.~G.~Lewin\altaffilmark{5}, Scott~F.~Anderson\altaffilmark{6}, David~Pooley\altaffilmark{7}
} 
\altaffiltext{1}{Institute of Astronomy $\&$ Department of Physics, National Tsing Hua University, Taiwan}
\altaffiltext{2}{SRON, Netherlands Institute for Space Research
 Sorbonnelaan 2, 3584 CA, Utrecht, The Netherlands}
\altaffiltext{3}{Department of Astrophysics, IMAPP, Radboud University Nijmegen
 PO Box 9010, 6500 GL Nijmegen, The Netherlands}
\altaffiltext{4}{Astronomical Institute, Utrecht University, P.O. Box 
80000, 3508 TA, Utrecht, the Netherlands}
\altaffiltext{5}{Kavli Institute for Astrophysics and Space Research,
Massachusetts Institute of Technology, 77
Massachusetts Avenue, Cambridge, MA 02139}
\altaffiltext{6}{Department of Astronomy, University of Washington, Box 
351580, Seattle, WA 98195}
\altaffiltext{7}{Astronomy Department, University of Wisconsin-Madison, 475 North
Charter Street, Madison, WI 53706}
\altaffiltext{8}{Kenda Foundation Golden Jade Fellow}

\begin{abstract}
We study a \textit{Chandra X-ray Observatory} ACIS-S observation of the 
Galactic globular cluster M12. With a 26 ks exposure time, we detect 
6 X-ray sources inside the half-mass radius (2$\farcm$16) of which two 
are inside the core radius (0$\farcm$72) of the cluster. If we assume these 
sources are all associated with globular cluster M12, the luminosity $L_\mathrm{X}$ 
among these sources between 0.3-7.0 keV varies roughly from $10^{30}$ to $10^{32}$ \,ergs\,s$^{-1}$. 
For identification, we also analyzed the \textit{Hubble Space Telescope}  (\textit{HST}) 
Advanced Camera for Surveys (ACS) and Wide Field and Planetary Camera 2 (WFPC2) data and identified the optical counterparts to five 
X-ray sources inside the \textit{HST} ACS field of view. According to the X-ray 
and optical features, we found 2-5 candidate active 
binaries (ABs) or cataclysmic variables (CVs) and 0-3 background galaxies within the \textit{HST} 
ACS field of view. Based on the assumption that the number of X-ray sources scales with 
the encounter rate and the mass of the globular cluster, we expect 2 X-ray source inside M12, and the expectation is consistent with our observational results.
Therefore, the existence of identified X-ray sources (possible CVs or ABs) in M12 suggests the primordial origin of 
X-ray sources in globular clusters which is in agreement with previous studies.
\end{abstract}


\section{Introduction}
Globular clusters are well known to be very efficient factories to produce exotic binary systems. 
The probability to find an X-ray source containing a compact object with luminosity $L_\mathrm{X} \gtrsim 10^{36}$ \,ergs\,s$^{-1}$
in a globular cluster is about two orders of magnitude larger than that in the Galactic disk (Katz 1975; Clark 1975). 
Theoretical studies suggest that those bright X-ray sources are binary systems with a neutron star or 
black hole and are formed via tidal capture or exchange companions with encounters (Fabian et al. 1975; Hills 1976). 
As for the low luminosity X-ray sources ($L_\mathrm{X}$ $\lesssim$ $10^{34.5}$ \,ergs\,s$^{-1}$), 
the population was first investigated by Hertz \& Grindlay (1983). 
These dim sources consist of several types of X-ray binary systems. The brightest are quiescent low-mass X-ray 
binaries (qLMXBs, an accreting neutron star or black hole with a companion star at a low accretion rate), and then 
cataclysmic variables (CVs, an accreting white dwarf with a main-sequence or subgiant companion), followed by 
active binaries (ABs, two main-sequence/subgiant objects in a binary), and millisecond pulsars (MSPs) which are thought 
to be the descendants of the LMXBs with their luminosity similar to CVs. Studying the population of X-ray sources in globular 
clusters can help us understand more about the physical properties of X-ray binaries and their formation mechanism, and construct 
the dynamical evolution scenario of globular clusters (Hut et al. 1992).

With the high resolution (down to the scale of sub-arcsecond) and better sensitivity of the \chandra\ X-ray
Observatory and 
\textit{Hubble Space Telescope} (\textit{HST}), some faint X-ray sources, especially CVs and ABs in globular clusters 
have been studied and identified, such as 47Tuc (Grindlay et al. 2001a; Edmonds et al. 2003; Heinke et al. 2005), NGC6397 (Grindlay et al. 2001b), 
$\omega$ centauri (Rutledge et al. 2002), NGC6752 (Pooley et al. 2002a), NGC6440 (Pooley et al. 2002b), NGC6626 (Becker et al. 2003), M80 (Heinke et al. 2003c), 
M4 (Bassa et al. 2004), NGC288 (Kong et al. 2006), NGC2808 (Servillat et al. 2008), M55 and NGC6366 (Bassa et al. 2008). 
These sources can be formed by primordial binaries (for the majority of active binaries and 
some cataclysmic variables) or dynamical interactions (for some cataclysmic variables). Pooley \& Hut (2006) and 
Bassa et al. (2008) report that the number of X-ray sources with luminosity $L_\mathrm{X} \lesssim 10^{34.5}$ \,ergs\,s$^{-1}$ in 
the 0.5-6.0 keV range will scale with the encounter rate (for the case of dynamical origin) and the total mass 
(for the case of primordial origin) of globular clusters. M12 has a relatively low encounter rate and is less massive  
when compared with many of the globular clusters previously studied.   
M12, as a low-density core globular cluster, 
is a good target to study the correlation between the number of X-ray sources 
and the encounter rate as well as the mass of globular clusters at the low core density end.
Based on previous studies of other globular clusters, more active binaries are identified in low-density core globular clusters than in dense-core globular clusters (e.g. M4; Bassa et al. 2004). 
So we could expect there will be more active binaries in M12. 
Furthermore, Verbunt (2002) indicate that magnetically active binaries (mainly RS CVn systems) are more likely to be primordial in origin. 
Hence, we could also expect that the number of X-ray sources has a stronger dependence on the total mass than on the encounter rate for the low-density core globular clusters. 

M12 is a relatively low-density core globular cluster with 0$\farcm$72 of core radius and 2$\farcm$16 
of half-mass radius (Harris 1996, version of 2003 February). It is near the galactic disk ($l=15\fdg72$, $b=26\fdg31$) 
so there are not so many (foreground or background) stars in the region as on the galactic disk. The absolute visual magnitude of M12 is -7.32. 
The distance to M12 is 4.9 kpc and 
the extinction toward M12 is E[B-V]=0.19 corresponding to a neutral hydrogen column density N$_\mathrm{H}$=1.0$\times10^{21}$ \,cm$^{-2}$ 
(derived from N$_\mathrm{H}$=5.3$\times$10$^{21}$E[B-V] \,cm$^{-2}$ by Predehl \& Schmitt 1995). We used these values in the following analysis. 

According to a previous optical study, M12 is an (optical) variable-poor cluster (Malakhova et al. 1997). 
Until 2001, only a few optical variables have been identified 
to be a W Vir variable (Sawyer 1938; Clement et al. 1988) or a W UMa variable (von Braun et al. 2002). 
In addition, some UV-bright stars have been discovered in the past decades (e.g. Zinn et al. 1972; Harris et al. 1983; Geffert et al. 1991).
Previous observation of M12 with HEAO-1 and Einstein (Hertz \&\ Wood 1985) and with ROSAT in its All Sky Survey (Verbunt et al. 1995) did not detect a source. The ROSAT upper limit is the lowest: $f_{0.5-2.5\mathrm{keV}}<1.7\times 10^{-13}$ ergs\,s$^{-1}$cm$^{-2}$.

In $\S2$ we describe the observation and analysis of  \chandra\ X-ray data. The \hst\ optical data is presented in $\S3$. 
In $\S4$ We discuss the source identification and in $\S5$ compare our results with other globular clusters.
\\
\begin{inlinefigure}
\medskip
\psfig{file=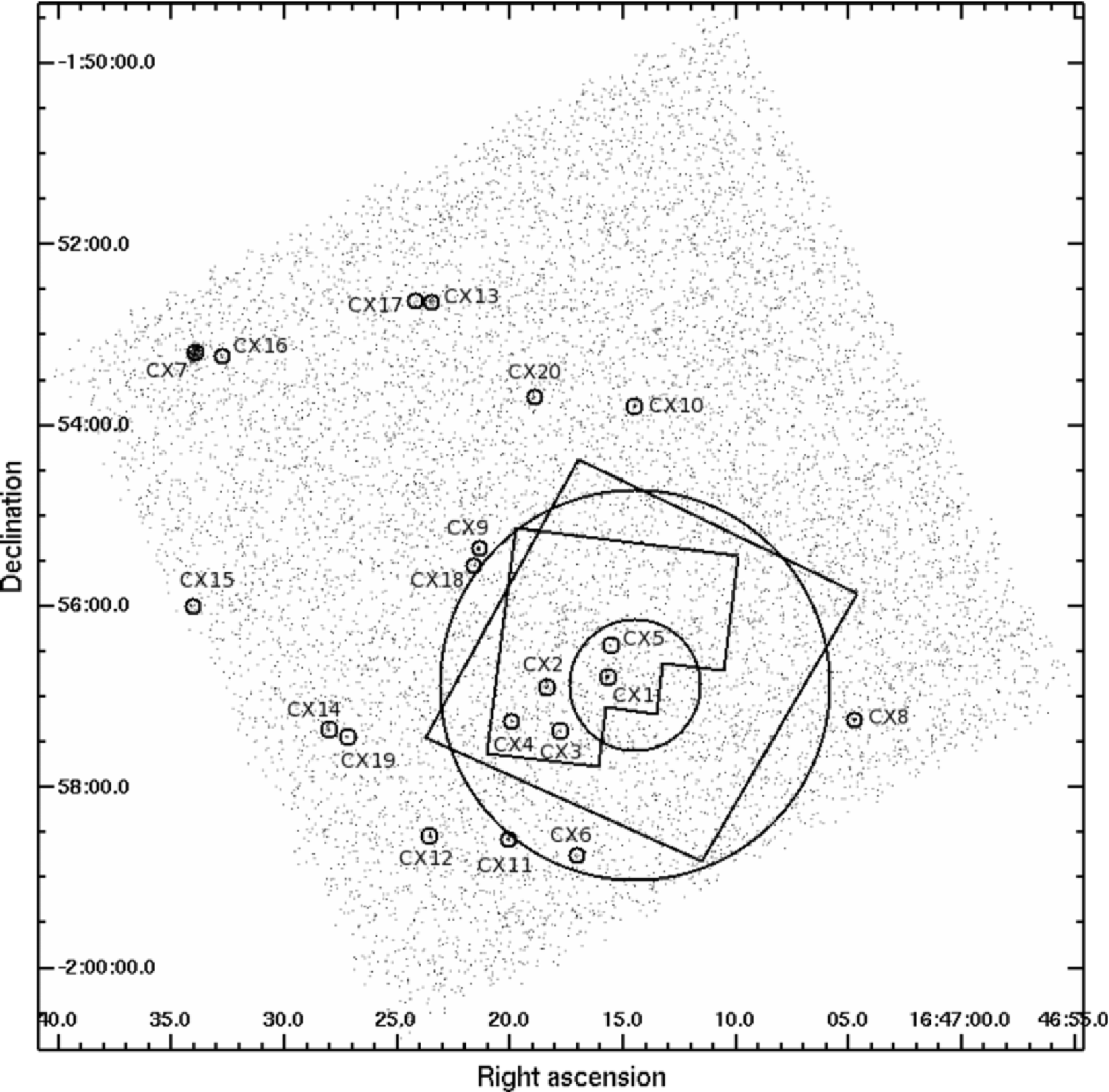,width=3.6in}
\centering
\caption{\chandra\ ACIS-S3 0.3-7 keV image of NGC6218. The 20 detected X-ray sources are marked and named with their CX number. 
The inner circle is the core radius, and the large circle is the half-mass radius of the globular cluster M12. The box is the field 
of view of the \textit{HST} ACS drizzled image. The polygon is the field of view of the \textit{HST} WFPC2 drizzled image.
}
\end{inlinefigure}

\section{X-ray Observations and Analysis}
M12 was observed on 2004 July 17 for 26.6 ks with the Advanced CCD Imaging Spectrometer (ACIS) on the \chandra\ X-Ray 
Observatory (ObsID 4530). The telescope aim point is on the ACIS back-illuminated S3 chip. The data were taken in the timed 
mode with a frame transfer time of 3.24 s and were telemetered to the ground in faint mode. The field of view of ACIS S3 chip 
($\sim$ 8.3 $\times$ 8.3  arcmin$^{2}$)\footnote{See The Chandra Proposers' Observatory Guide at http://cxc.harvard.edu/proposer/POG/html/index.html} 
covering the whole half-mass radius of M12. In this paper, we only consider the data taken with the S3 chip.

\subsection{Data Reduction}
We used \textsc{CIAO}, version 3.4 provided by the \chandra\ X-Ray Center (CXC)\footnote{See http://cxc.harvard.edu/} to perform data reduction and analysis. 
We reprocessed the level 1 event files with \textsc{CALDB}, version 3.4.2. The reprocessing included flagging and filtering out the cosmic rays and creating a new 
bad pixel file by using \texttt{acis$\_$run$\_$hotpix} package. The data were also filtered with the \texttt{ASCA} (Advanced Satellite for Cosmology and Astrophysics) grades 
of 0, 2, 3, 4, and 6 which is the standard value to optimize the instrumental signal-to-background ratio. We only processed the events extracted with photon 
energies in the range of 0.3-7 keV. Periods of high background flares 
(count rate $>$ 5 counts \,s$^{-1}$ which is inspected from the entire area of chip S1) were eliminated. The final effective exposure time was 26.5 ks.

\begin{table*}
\centering{\footnotesize
\caption{\chandra\ Source Properties}
\begin{tabular}{lccccccccc}
\hline
\hline
Source & R.A. & Decl. &\multicolumn{3}{c}{Net Counts} & $f_\mathrm{0.3-7}$& $f_\mathrm{0.5-2.5}$ \\
 \cline{4-6}
Name  & (J2000.0) & (J2000.0) & Soft & Medium & Hard & & & Counterpart \\
\hline
CX1	 & 16:47:15.687 (0.02) &	-01:56:46.90 (0.03) &	55.9 &	124.9 &	77.7 &	7.36E-14 &	3.62E-14 & \hst\ \\
CX2 & 16:47:18.396 (0.19) &	-01:56:53.62 (0.10) &	2.0 &	6.6 &	4.9 &	3.84E-15 &	1.89E-15 & \hst\ , WFI\\
CX3	 & 16:47:17.776 (0.16) &	-01:57:22.68 (0.15) &	0.8 &	4.0 &	3.0 &	2.22E-15 &	8.98E-16 & \hst\ \\
CX4 & 16:47:19.906 (0.20) &	-01:57:16.32 (0.16) &	4.7 &	1.0 &	1.0 &	1.90E-15 &	9.92E-16 & \hst\ , WFI\\
CX5	 & 16:47:15.534 (0.16) &	-01:56:25.89 (0.16) &	1.0 &	2.0 &	2.0 &	1.43E-15 &	5.19E-16 & \hst\ \\
CX6 & 16:47:17.044 (0.26) &	-01:58:45.52 (0.14) &	1.0 &	4.0 &	 0.1 &	1.66E-15 &	8.98E-16 & WFI\\
\\
CX7 &	16:47:33.944 (0.13) &	-01:53:11.78 (0.21) &	41.5 &	106.4 &	79.5 &	6.47E-14 &	2.97E-14 \\
CX8 & 16:47:04.745 (0.06) &	-01:57:15.25 (0.04) &	13.0 &	37.0 &	27.7 &	2.21E-14 &	9.34E-15\\
CX9 &	16:47:21.384 (0.08) &	-01:55:21.41 (0.09) &	11.0 &	28.0 &	11.0 &	1.43E-14 &	7.55E-15 \\
CX10	 & 16:47:14.475 (0.12) &	-01:53:47.01 (0.18) &	9.5 &	14.9 &	3.8 &	8.00E-15 &	3.91E-15 \\
CX11 &	16:47:20.082 (0.10) &	-01:58:34.42 (0.10) &	1.0 &	13.0 &	13.0 &	7.69E-15 &	2.87E-15 &WFI\\
CX12 &	16:47:23.558 (0.09) &	-01:58:32.32 (0.08) &	10.0 &	8.9 &	2.8 &	6.18E-15 &	3.53E-15 &WFI\\
CX13 &	16:47:23.471 (0.31) &	-01:52:37.86 (0.22) &	3.6 &	6.3 &	5.5 &	4.39E-15 &	1.92E-15 \\
CX14 &	16:47:28.032 (0.12) &	-01:57:21.97 (0.23) &	 0.1 &	0.0 &	11.5 &	3.77E-15 &	3.13E-16 \\
CX15 &	16:47:34.039 (0.25) &	-01:56:00:04 (0.22) &	0.8 &	4.8 &	2.4 &	2.27E-15 &	9.63E-16 \\
CX16 &	16:47:32.744 (0.33) &	-01:53:14.12 (0.27) &	2.5 &	1.0 &	4.3 &	2.23E-15 &	9.95E-16 \\
CX17 &	16:47:24.167 (0.32) &	-01:52:37.44 (0.33) &	2.8 &	2.3 &	1.8 &	1.95E-15 &	1.27E-15 \\
CX18 &	16:47:21.574 (0.07) &	-01:55:32.86 (0.14) &	1.0 &	2.8 &	2.9 &	1.90E-15 &	8.63E-16 \\
CX19 &	16:47:27.191 (0.30) &	-01:57:26.42 (0.11) &	1.0 &	0.0 &	3.0 &	1.14E-15 &	0.00E+00 \\
CX20 & 16:47:18.899 (0.13) &	-01:53:41.13 (0.13) &	 0.2 & 1.7 &	0.6 &	7.85E-16 &	4.13E-17 \\
\hline
\hline
\end{tabular}
}
\par
\medskip
\begin{minipage}{0.95\linewidth}
NOTES. --- 
We have performed astrometry correction for the positions determined by \texttt{wavdetect} listed in table 1. 
The R.A. and Decl. was corrected by -0$\farcs$05 and -0$\farcs$09 respectively and then corrected by the boresight correction (See $\S3.3$). 
The position uncertainties 
are given by \texttt{wavdetect} and in units of arcseconds. Units of right ascension are hours, minutes, and 
seconds, and units of declination are degrees, arcminutes, and arcseconds (J2000.0). The unabsorbed flux is 
derived assuming a power-law model with N$_\mathrm{H}$ = $10^{21}$ \,cm$^{-2}$ and a photon index of 2. 
The sources listed in the upper part are those inside the half-mass radius and the lower ones are outside the half-mass radius of M12.
\end{minipage}
\end{table*}

\subsection{Source Detection}
The \textsc{CIAO} package \texttt{wavdetect} was employed to detect X-ray sources inside the ACIS-S3 chip. 
The wavelet scales (also the sizes of the point spread function, PSF) were set to be a series from 1 to 16 
increasing by a factor of  $\sqrt{2}$. We also set the detection signal threshold to be $10^{-6}$ such that 
there will be at most one false detection (fake source) caused by background fluctuation. The data was 
divided into three energy bands: soft band (0.3-1 keV), medium band (1-2 keV), and hard band (2-7 keV). 
We performed \texttt{wavdetect} with exposure maps on the total energy band (0.3-7 keV) and also the three sub-bands with the parameters 
described above. We then produced a master source list 
which was the combined source list of the 4 energy bands. We detected 20 sources in the whole ACIS-S3 
chip and 6 of them are inside the half-mass radius of M12. Figure 1 shows the resultant source detection on the chip.
Table 1 shows the information of the 20 sources ordered by increasing unabsorbed flux within 0.3-7 keV. 
The columns list the source name, the position of the source (J2000.0), the net counts (source counts excluding the background contribution) 
in the three bands ($X_\mathrm{soft}$, $X_\mathrm{medium}$, and $X_\mathrm{hard}$), and the unabsorbed flux within 0.3-7 keV and 0.5-2.5 keV. 
The photon counts from the three bands were extracted from the source region corresponding to the 3$\sigma$ \texttt{wavdetect} ellipse. Furthermore, the background counts were extracted from an annulus centered at the individual source and outside the 3$\sigma$ \texttt{wavdetect} ellipse region. 
The unabsorbed flux was calculated from the source counts rate by assuming a power law model with 
a column density N$_\mathrm{H}$ = $10^{21}$ \,cm$^{-2}$ and a photon index of 2. 

We also estimated the possible number of background sources. If we consider the 14 sources outside the cluster half-mass radius as background sources, 
then we can scale the background sources inside the half-mass radius by the area ratio. Therefore, there will be 
14 $\times$ $\pi$r$_\mathrm{h}^{2}$ / ($\pi$r$_\mathrm{S3}^{2}$ - $\pi$r$_\mathrm{h}^{2}$) $\approx$ 
5 background sources inside the half-mass radius of M12. This is consistent with the log N - log S relation derived from the \chandra\ Deep Field Survey (Giacconi et al. 2001). 
Sources detected in this survey are active galactic nuclei (AGNs) and distant galaxies, 
and therefore we took them as background sources unrelated to the galactic globular cluster. 
Based on the log N - log S relation and the detection limit of the observation, we expected 16 to 22 out of 20 within the ACIS-S3 chip 
and 3 to 5 out of 6 within the half-mass radius to be background (AGNs) contribution. 
For an expectation of 5 backgrounds and by a Poisson distribution, the probability of detecting 6 or more backgrounds is roughly 38$\%$. 
If we use the completeness limit of the observation to estimate the backgrounds, we would have 3 to 4 backgrounds out of 6 sources inside the half-mass radius, which is consistent with the estimation using the detection limit. For the case of adopting completeness limit, the Poisson probability that all of the sources inside the half-mass radius are backgrounds is $\sim$ 21$\%$, and the result does not change our conclusion.
Therefore we cannot exclude the possibility that all our sources inside the half-mass radius are backgrounds unrelated to M12.

\begin{figure*}
\psfig{file=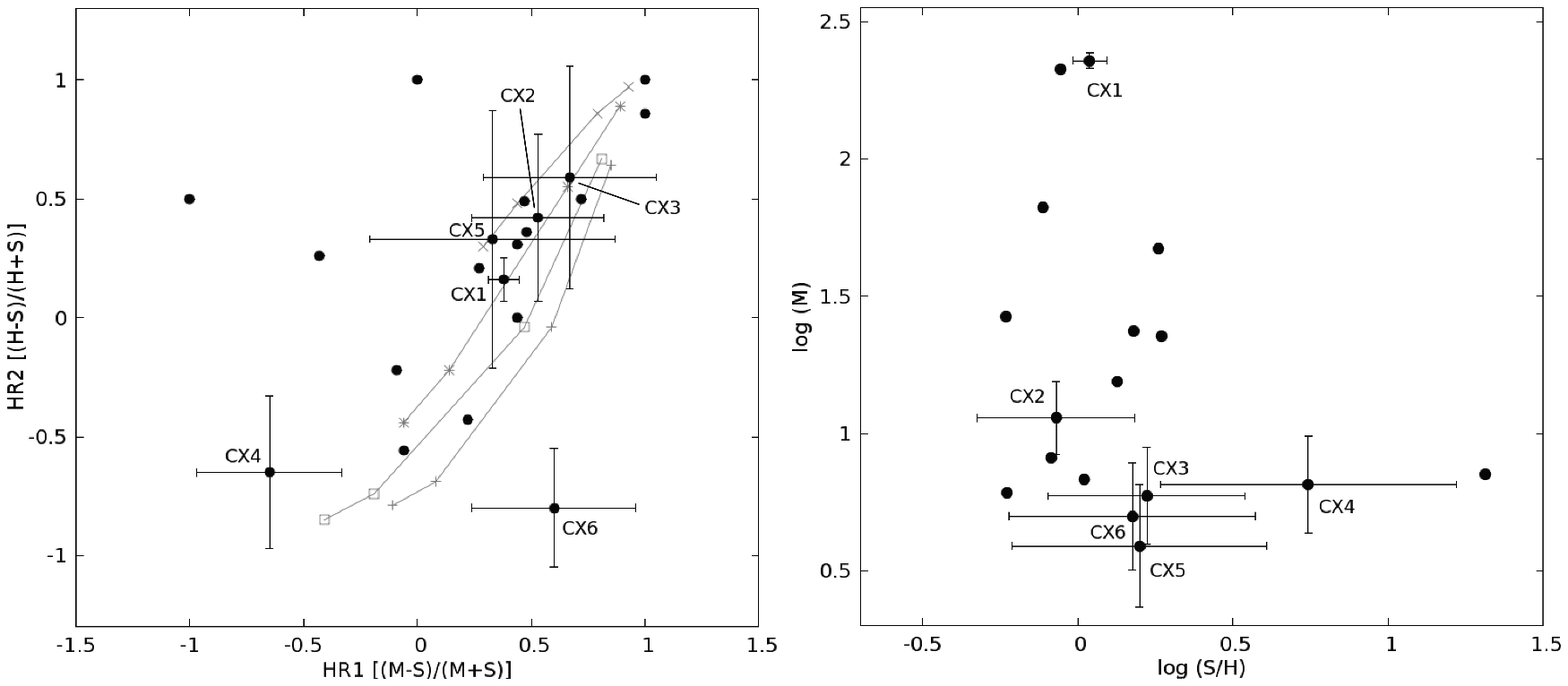,width=7.25in}
\centering
\caption{\textbf{Left} : The color-color diagram of the X-ray sources in M12. The number represents the source number within 
the half-mass radius. The lines are the hardness ratios predicted from thermal bremsstrahlung (TB) model and power 
law (PL) model with different column densities. From top to bottom are PL with a photon index of 1, PL with a photon index of 2, PL with a photon index of 3, 
and TB with temperature 1keV. The column density from left to right is 1 $\times$ $10^{20}$, 1 $\times$ $10^{21}$, 5 $\times$ $10^{21}$, and 1 $\times$ $10^{22}$ \,cm$^{-2}$.
\textbf{Right} : The color-magnitude diagram of the X-ray sources in M12. The X-ray color is defined as the logarithm of the ratio of 0.5-1.5 keV (S) counts to 1.5-6 keV (H) counts, 
and the magnitude is the logarithm of 0.5-4.5 keV (M) counts. Sources inside the half-mass radius are marked with error bars. 
}
\end{figure*}

\begin{inlinetable}
\caption{Spectral Fits of the Brightest Sources}
\begin{tabular}{lccccc}
\hline 
\hline
CX & Model\tablenotemark{a} & N$_\mathrm{H}$\tablenotemark{b} & kT/$\alpha$ & $\chi^2_\mathrm{\nu}/dof$ & $f_\mathrm{0.3-7}$\tablenotemark{c}\\
\hline
CX1 & TB & 1.78$\pm$0.90 & 3.08$\pm$1.18 & 1.67/12 & 5.17\\
  & PL & 3.14$\pm1.27$& 2.25$\pm$0.33 & 1.55/12 & 5.32\\
CX7 & TB & 1.65$\pm$1.05 & 12.14$\pm$12.65 & 1.53/11 &  6.84\\
  & PL & 2.07$\pm$1.07 & 1.53$\pm$0.30 & 1.53/11 & 6.98\\
\hline
\hline
\end{tabular}
\par
\medskip
\begin{minipage}{0.82\linewidth}
NOTES. --- All quoted uncertainties are 1$\sigma$.\\
$^a$ TB: thermal bremsstrahlung; PL: power-law.\\
$^b$ in units of $10^{21}$ \,cm$^{-2}$\\
$^c$ 0.3--7 keV unabsorbed flux in units of $10^{-14}$\flux.\
\end{minipage}
\end{inlinetable}

\subsection{X-ray Colors and Spectral Fitting}

Because most of our detected X-ray sources are faint, the photon counts are too few for the 
statistics to provide useful information from the X-ray spectrum. However, we can make a rough estimate of 
the X-ray properties from the hardness ratio constructed from the photon counts 
in different energy bands.  We used the CIAO tool 
\texttt{dmextract} to extract the photon counts, and the hardness ratio is defined 
as HR1 = $(X_\mathrm{medium} - X_\mathrm{soft}) / (X_\mathrm{medium} + X_\mathrm{soft})$  and  HR2 = $(X_\mathrm{hard} - X_\mathrm{soft}) / (X_\mathrm{hard} + X_\mathrm{soft})$. 
The $X_\mathrm{soft}$, $X_\mathrm{medium}$, and $X_\mathrm{hard}$ represent the photon counts extracted from the three energy ranges,  
0.3-1, 1-2, and 2-7 keV. Figure 2 shows the X-ray color-color diagram (left panel) and the color-magnitude 
diagram (right panel) of all the 20 sources detected on the ACIS-S3 chip. The four lines on the 
color-color diagram are the spectra predicted by thermal bremsstrahlung model and power law model 
with different column densities. Moreover, we made the energy ranges 
of the color-magnitude diagram to be consistent with previous works (e.g. Pooley et al. 2002a$\&$b, Heinke et al. 2003a, Bassa et al. 2004, Kong et al. 2006).

We extracted the spectrum for the sources CX1 and CX7 whose photon counts $\gtrsim$ 200 in 0.3-7 keV energy band. 
The counts per spectral bin is 15 and we fitted them with a power law model and thermal bremsstrahlung model. 
We fixed the Galactic column density N$_\mathrm{H}$ at $10^{21}$ \,cm$^{-2}$ derived from optical extinction for the fitting. All the models 
for the two sources can be fitted with a reduced $\chi^{2}$ $\sim$ 1.5 and the null hypothesis probabiility is $\sim$ 0.1. 
The null hypothesis probability is the probability of getting a value of $\chi^{2}$ as large or larger than observed if the model is correct. 
If this probability is small then the model is not a good fit.
Table 2 summarizes the results of spectral fitting. 
We also allowed for an additional intrinsic column density to the source beyond the Galactic value, and in each case this 
fitted intrinsic absorbing is higher than the (fixed) Galactic value. The temperature for thermal bremsstrahlung model is 
3 keV for CX1 and 12 keV for CX7. The photon index for CX1 is roughly 2 while for CX7 it is slightly lower 
than 2. Besides, the predicted unabsorbed flux (0.3-7.0 keV) is consistent with each other in both models. We also 
extracted the light curve (0.3-7.0 keV) of CX1, with a time resolusion of 2000 seconds. 
We then performed a Kolmogorov-Smirnov (K-S) test on the light curve of CX1 by using
LCSTATS in the XRONOS (version 5.21) package. 
The probability of constancy is 4.79 $\times$ $10^{-3}$.
Figure 3 shows the power law fitted spectrum and the light curve of CX1.

\begin{figure*}
\psfig{file=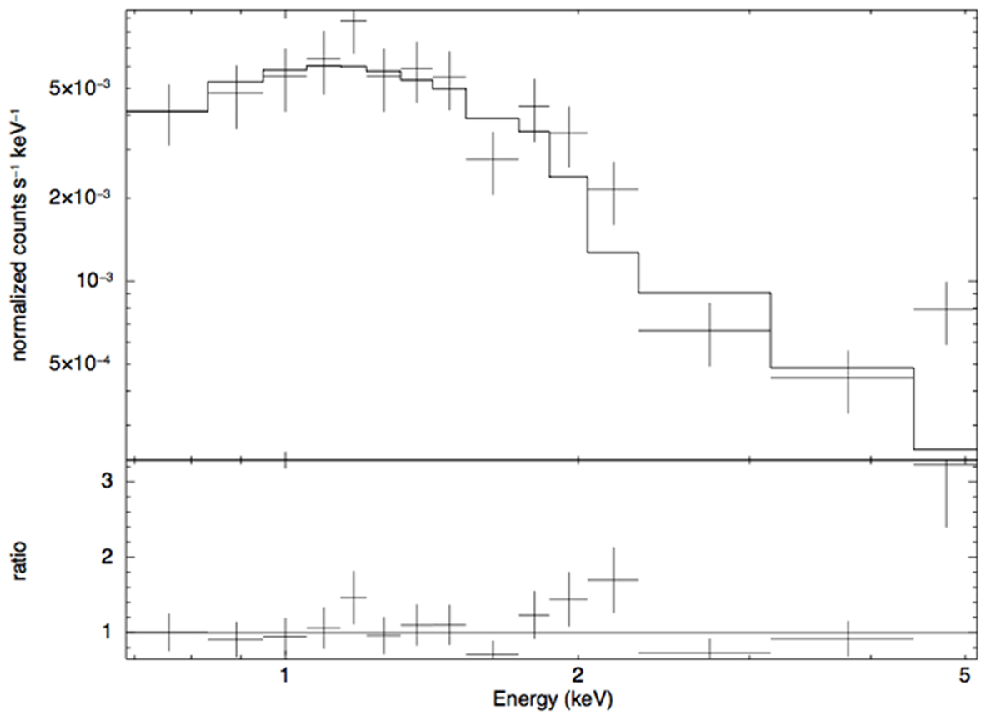,width=3.55in}
\psfig{file=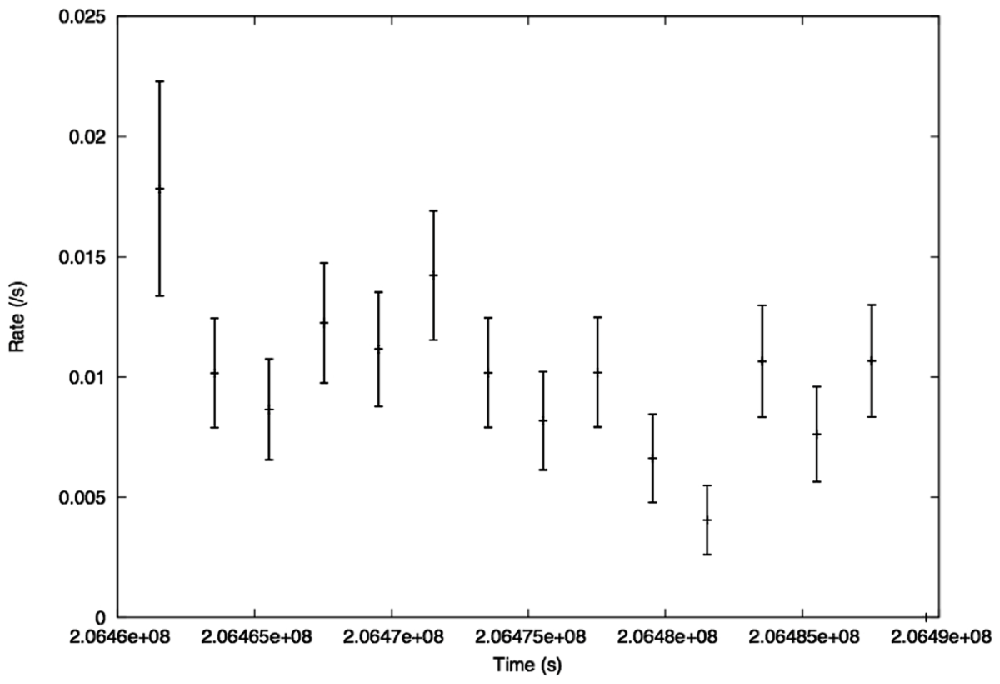,width=3.75in}
\caption{\textbf{Left}:The spectrum of the most brightest source (CX1) within the core radius of M12. 
The cross represents the data points. The solid line is the fitted absorbed power-law model. The lower panel is the ratio.
The fitted photon index and N$_\mathrm{H}$ is 2.2 $\pm$ 0.3 and 3 $\pm$ 1 $\times$ $10^{21}$ \,cm$^{-2}$ respectively.  
The null hypothesis probability = 1.002 $\times$ $10^{-1}$. \textbf{Right}:The light curve of CX1 in the energy range 0.3 - 7.0 keV. 
For each bin time is 2000 seconds. Total exposure time is $\sim$ 26.5 ks. The probability of constancy tested by a Kolmogorov-Smirnov (K-S) test is 4.79 $\times$ 10$^{-3}$.}
\end{figure*}

\section{Optical Observations}
M12 was observed by \hst\ Advanced Camera for Surveys (ACS) on
 2004 June 14 (Proposal ID 10005). In this observation, three filters were used: F435W ($B_\mathrm{435}$), F625W ($r_\mathrm{625}$), 
 and F658W ($H\alpha_\mathrm{658}$). The exposure time for F435W, F658W, and F625W is 1360 seconds, 1360 seconds, and 200 seconds, respectively. 
 The field of view of ACS covers the whole core radius of M12 but only covers $\sim$75\% of the half-mass radius. 
 There are five X-ray sources within the ACS field of view. The five sources are also covered by the \hst\ Wide Field and Planetary Camera 2 (WFPC2) observation (Proposal ID 8118) (See figure 1). The WFPC2 observation was performed on 2000 August 12. Two filters F439W and F555W were employed in the observation, and the exposure time for F439W and F555W is 240 seconds and 63 seconds respectively.

\subsection{Data Reductions}
We used individual images for photometry and the ACS drizzled images for astrometry 
and identifying the optical counterparts for X-ray sources. The drizzled images are combined images which 
have been calibrated for bias, dark, and flat field. The geometric distortion and cosmic rays are also removed. 
The calibrations were performed by ACS calibration pipeline with the tool 
PyDrizzle\footnote{See the \textit{HST} ACS Data Handbook at http://www.stsci.edu/hst/acs/documents/handbooks/DataHandbookv2/ACS$_{-}$longdhbcover.html}.

As for the WFPC2 images, we used them to perform photometry only. The calibrated images were adopted in the data analysis. The bias and flatfield corrections were applied on the calibrated images by WFPC2 calibration pipeline\footnote{See the \hst\ WFPC2 Data Handbook at http://www.stsci.edu/instruments/wfpc2/Wfpc2$_{-}$dhb/WFPC2$_{-}$longdhbcover.html}.

\subsection{Photometry}
For photometry, we used individual images and processed them with the package \textsc{DOLPHOT},  a PSF photometry tool adapted from 
\textsl{HSTphot} (Dolphin 2000)\footnote{See http://purcell.as.arizona.edu/dolphot/}. \textsc{DOLPHOT} can process multiple images at the same field of view 
for one run and provide the results of combined photometry for each filter. We applied the \textsc{DOLPHOT} with ACS module in all analysis. 
First, we run the \texttt{acsmask} command on the data quality images provided by Space Telescope Science Institute (STScI) in order to mask bad pixels. 
Then we created a sky map by applying the \texttt{calcsky} command. In the last step, 
we performed \texttt{dolphot} with ACS PSF and Pixel Area Maps provided by the \textsc{DOLPHOT} ACS module. A master photometry list was given containing the position, magnitude for each bands, signal-to-noise ratio, and other indicators for the detected stars. We set a criteria to eliminate the cosmic rays, 
artifacts, and the fake stars located on the diffraction spikes of the saturated stars. 
Finally we filtered the output stars and got a $\arcsec$good$\arcsec$ star list. Stars were 
selected if they showed up in all three bands. 
Then we produced the color-magnitude diagrams (CMDs) with the $\arcsec$good$\arcsec$ star list (See figure 4, left two diagrams). 

We used processes similar to that of  \hst\ ACS data analysis as we performed photometry on the \hst\ WFPC2 data, while we employed the package \textsl{HSTphot} (version 1.1) instead of \textsc{DOLPHOT}. We first applied the \texttt{mask} command to mask bad pixels, and then calculated the sky map by running \texttt{getsky}. The task \texttt{crmask} was employed to filter out the cosmic rays on the images. We also removed the hot pixels with the task \texttt{hotpixels}. Finally we combined the image frames for each filter into a single image by the command \texttt{coadd}, and then performed the task \texttt{hstphot} on the resultant single image for each filter to calculate the photometry. \textsl{Hstphot} can deal with multiple images with different filters simultaneously and produce a master list containing the photometry information for each filter. As a consequence, we produced the CMD of the WFPC2 observation (See figure 4, right diagram).

We further utilized the package \textsl{DAOPHOT} (Stetson 1987) in \textsl{IRAF}\footnote{See the user guide of IRAF/DAOPHOT at http://iraf.noao.edu/iraf/ftp/iraf/docs/daorefman.ps.Z} (version 2.12.2a) to extract the photometry information of the possible optical counterpart of CX1 from WFPC2 observation. The CX1 counterpart is located on the edge of WFPC2 chip 3 (WFC3), where the image quality might not be as good as that in the central region of the CCD. We adopted relative photometry using the \textsl{HSTphot} photometry as reference. We then calibrated the photometry by shifting the stars' magnitudes extracted by \textsl{DAOPHOT} to match the same stars' magnitudes extracted by \textsl{HSTphot}. Finally, we applied the magnitude shifts on the CX1 counterpart and plotted it on the CMD (See figure 4, right diagram).

\begin{figure*}
\psfig{file=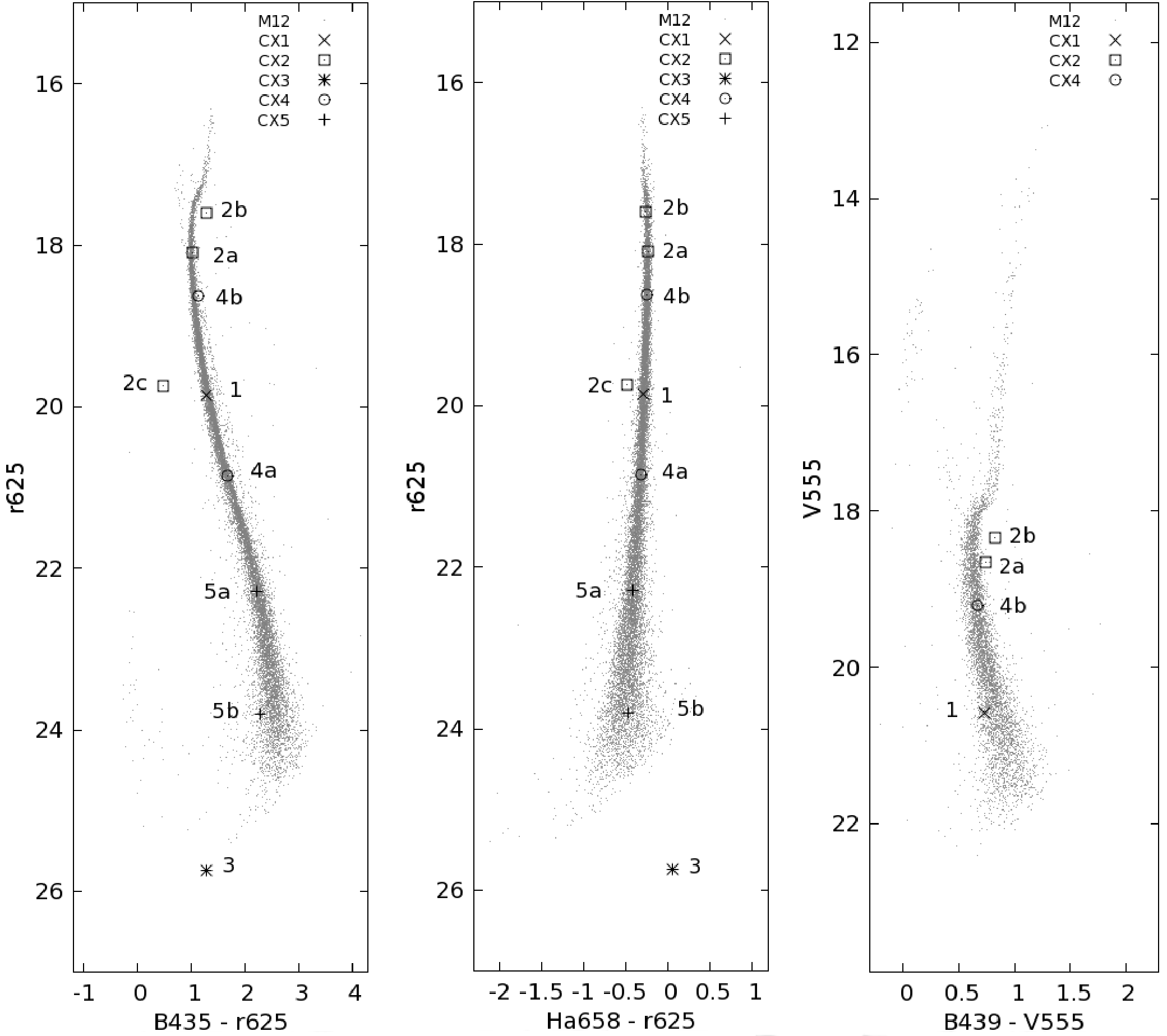,width=7.2in}
\centering
\caption{The left two diagrams are the color-magnitude diagrams (CMDs) of \textit{HST} ACS observation of M12. The right diagram is the CMD of \hst\ WFPC2 observation of M12. The number represents the X-ray source number within the half-mass radius.}
\end{figure*}

\begin{figure*}
\psfig{file=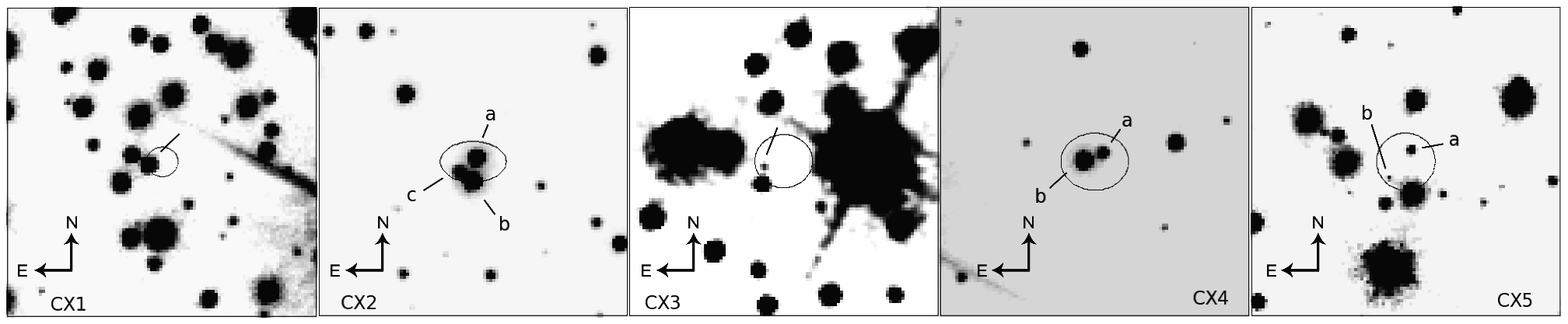,width=7.2in}
\centering
\caption{5$\arcsec$ $\times$ 5$\arcsec$ finding charts for the X-ray sources inside the \hst\ ACS field of view. The background optical image is the drizzled R-band (F625W) \hst\ image. The 95$\%$ error circles of X-ray sources  have been laid on the charts. The optical counterpart candidates are marked with letters. The gray scales of the five images are different in order to show the counterparts clearly.}
\end{figure*}

\subsection{Astrometry}
In order to search for the optical counterparts of the \chandra\ X-ray sources, we placed the optical and the X-ray images on the 
same image frame and coordinate system. We took a wide field optical image as the reference image frame and aligned the \hst\ ACS and 
\chandra\ images onto it individually. The wide field image was retrieved from a V-band image of M12 taken with the ESO 2.2m telescope 
Wide Field Imager (WFI) on 2002 June 20 for 100 seconds. The WFI has 2 $\times$ 4 CCDs array, each CCD has 
8$\arcmin$ $\times$ 16$\arcmin$ field of view, resulting 
in a total  33$\arcmin$ $\times$ 34$\arcmin$ field of view. M12 was on one of the 8 chips (chip 7) with the whole area covering the half-mass radius. 
We first did the bias and flat-field calibrations, and then performed the astrometry on the image extracted from that chip only.

We chose 173 stars on the WFI image that matched the USNO CCD Astrograph Catalog (UCAC2) standards 
(Zacharias et al. 2004) and then fitted them with a six-parameter transformation. As a result, we obtained a transformation solution with 
residual errors of 0$\farcs$087 in R.A. and 0$\farcs$093 in Decl. 
We then chose 317 stars on the \hst\ ACS drizzled F625W image and matched them with the WFI image. 
The resultant residual error of the transformation solution was 0$\farcs$036 in R.A. and 0$\farcs$021 in Decl. 
At this stage, we have aligned the WFI image and the \hst\ ACS image together and placed them both on the absolute ICRS frame.

For the \chandra\ image, we first improved its absolute astrometry with the Aspect Calculator\footnote{See http://cxc.harvard.edu/ciao3.4/threads/arcsec$_{-}$correction/}
privided by the \chandra\ X-ray Center. 
It provides an absolute astrometry of 0$\farcs$6 (90\%). The calculated offset is -0$\farcs$05 in R.A. and -0$\farcs$09 in Dec. 
We found CX2, CX4, CX6, CX11 and CX12 coincided with possible counterparts in the WFI image when we overlaid the 99\% confidence error circles 
of \chandra\ sources on the WFI image. 
CX2 has 3 \hst\ optical counterparts, and in the WFI image the three counterparts are not resolved. CX4 has 2 optical counterparts on the \hst\ image. 
Since we can not be sure which source will be the correct counterpart, we did not take CX2 and CX4 as astrometry references.
In addition, CX11 corresponds to a non-point-like source while CX6 corresponds to a point-like source. 
Although CX6 corresponds to a WFI source just beside a saturated star, the WFI counterpart is bright enough and could be distinguished from the saturated star, 
and in addition, could be located precisely. 
Based on its brightness and position accuracy, we took it as an astrometry reference. 
The optical counterpart of CX12 on the WFI image is also a point-like source and outside the half-mass radius of M12. 
The stellar density outside the half-mass radius is so low that we consider the identification secure. 
As a consequence, we used the optical counterparts of CX6 and CX12 to perform boresight correction on \chandra\ image. 
The position offset needed to apply on the \chandra\ sources is 0$\farcs$421 $\pm$ 0$\farcs$040 in R.A. and -0$\farcs$042 $\pm$ 0$\farcs$004 in Decl. 
After the boresight correction, the WFI and the \chandra\ image are on the same frame. 

With the X-ray and optical astrometry, we obtained a final uncertainties of the X-ray source position in R.A. and in Decl., which are the root of 
the quadratic sum of the position uncertainty of the X-ray sources (given in Table 1), the uncertainty of the absolute astrometry in the WFI, 
the uncertainty of the astrometry in the WFI and \hst\ image alignment, 
and the uncertainty of the X-ray boresight correction. The final 1$\sigma$ position error for all X-ray sources varies roughly from 0$\farcs$1 to 0$\farcs$4. 
We took all the \textit{HST} sources inside the $95\%$ ($\sim$ 2.48$\sigma$) confidence X-ray error circle as the possible counterparts in optical band. Their finding charts are listed in figure 5.
\\ 
\\
\begin{inlinefigure}
\psfig{file=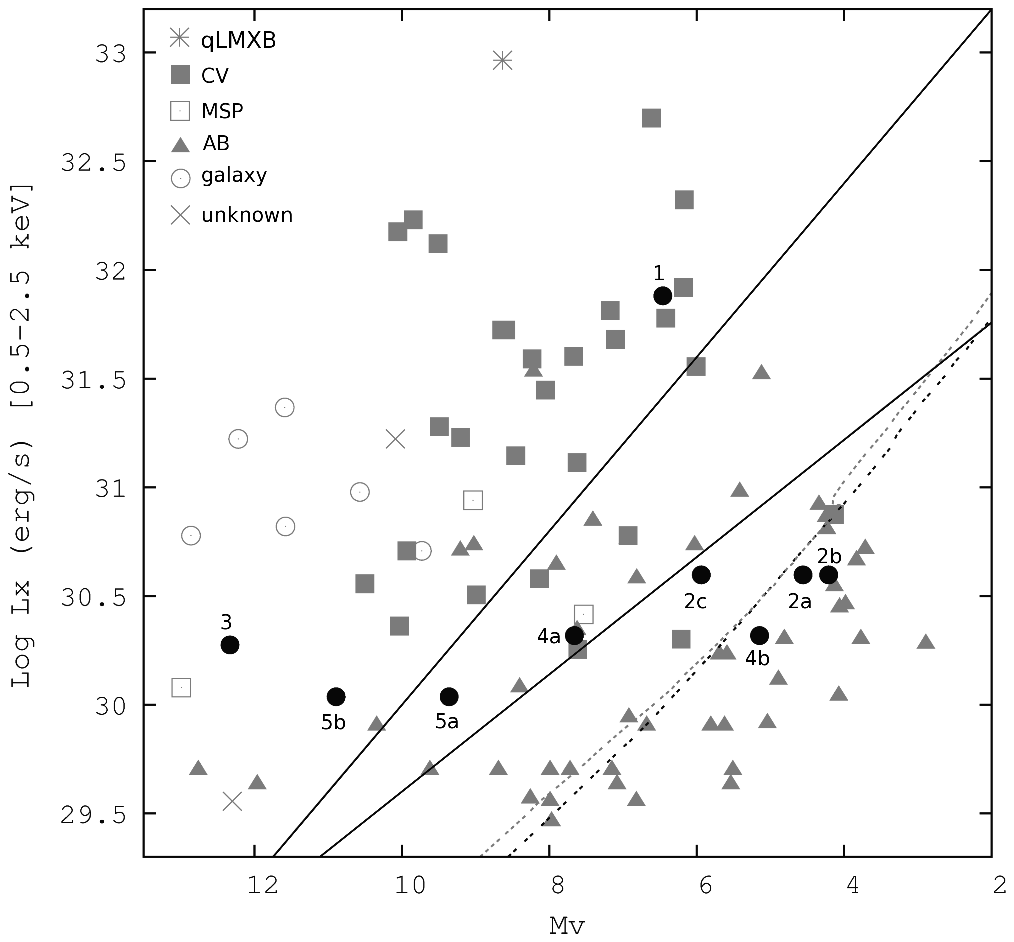,width=3.4in}
\centering
\caption{The diagram of the 0.5-2.5 keV X-ray  luminosity to the V-band absolute magnitude. 
The V-band magnitudes are derived from V = ($B_\mathrm{435}$ + $r_\mathrm{625}$)/2. The upper solid line represents the 
formula : Log $L_\mathrm{X}$ = 34.0 - 0.4$M_{V}$ (after Bassa et al. 2004) that is the constant X-ray to optical flux ratio roughly separating CVs and ABs. The lower solid line represents the formula : Log $L_\mathrm{X}$ = 32.3 - 0.27$M_{V}$ (after Verbunt et al. 2007 and Bassa et al. 2008) that gives an upper bound to the X-ray luminosities of stars and RS CVn binaries near the sun. The two dashed lines are computed from 11.2 Gyr isochrones for z = 0.019 (upper) and z = 0.001 (lower) from Girardi et al. (2000) and represent the maximum X-ray luminosities of ABs in the solar neighbourhood and in globular clusters respectively, 
assuming $L_\mathrm{X}$ $\simeq$ 0.001$L_\mathrm{bol}$.
The optical candidates are the solid dots marked with their names. Other data points are the sources identified from 47Tuc, M4, NGC6752, and NGC288.}
\end{inlinefigure} 
\\

\section{Source Identification and Classification}

\begin{sidewaystable*}
\centering{\footnotesize
\caption{Optical Counterparts to \chandra\ X-ray Sources}
\begin{tabular}{lccccccccccc}
\hline
\hline
 & $\Delta$ R.A. & $\Delta$ Decl. & & & & & & & &\\
CX & (arcsec) & (arcsec) & $B_\mathrm{435}$ & $B_\mathrm{439}$ & $V_\mathrm{555}$ & $r_\mathrm{625}$ & H$\alpha_\mathrm{658}$ & $L_\mathrm{X(0.3-7)}$\tablenotemark{b} & $M_\mathrm{V}$\tablenotemark{c} & $f_\mathrm{X}/f_\mathrm{r}$\tablenotemark{d} & Classification\tablenotemark{e}\\
\hline
1 &	0.20 &	0.06 &	21.15 $\pm$ 0.004 & 20.58 $\pm$ 0.16 & 21.31 $\pm$ 0.15 &	19.86 $\pm$ 0.006 &	19.57 $\pm$ 0.01 &	2.11E+32 & 6.46 & 3.151 & CV\\
2a &	-0.08 &	0.17 &	19.11 $\pm$ 0.001 & 19.39 $\pm$ 0.02 & 18.65 $\pm$ 0.01 &	18.09 $\pm$ 0.002 &	17.85 $\pm$ 0.004 &	1.10E+31 & 4.56 &	0.032 & chance counterpart \\
2b &	0.00 &	-0.20 &	18.88 $\pm$ 0.001 & 19.16 $\pm$ 0.06 & 18.34 $\pm$ 0.02 &	17.61 $\pm$ 0.002 &	17.35 $\pm$ 0.003 &	1.10E+31 & 4.21 &	0.021 & AB\\
2c &	0.20 &	-0.06 &	20.22 $\pm$ 0.003 &...&...&	19.75 $\pm$ 0.006 &	19.26 $\pm$ 0.01 & 1.10E+31 & 5.94 &	0.149  & CV\\
3 &	0.32 &	0.03 &	27.03 $\pm$ 0.16 &...&...& 25.74 $\pm$ 0.37 &	25.80 $\pm$ 0.94 &	6.37E+30 & 12.33 &	12.703 & chance counterpart \\
4a &	 -0.14 &	0.26 &	22.53 $\pm$ 0.008 &...&...&	20.86 $\pm$ 0.01 &	 20.54 $\pm$ 0.02 &	5.46E+30 & 7.66 &	0.205 & chance counterpart \\
4b &	0.17 &	0.13 &	19.75 $\pm$ 0.002 & 19.87 $\pm$ 0.02 & 19.20 $\pm$ 0.01 &	18.63 $\pm$ 0.003 &	18.39 $\pm$ 0.005  &	5.46E+30 & 5.15 &	0.026 & chance counterpart \\
5a &	 -0.11 &	0.29 &	24.51 $\pm$ 0.03 &...&...& 22.29 $\pm$ 0.02 &	21.87 $\pm$ 0.04 & 4.10E+30 & 9.36 &	0.573 & chance counterpart \\
5b &	0.26 &	-0.18 &	26.09 $\pm$ 0.08 &...&...& 23.77 $\pm$ 0.06 &	23.32 $\pm$ 0.12 & 4.10E+30 & 10.89 &	2.298 & chance counterpart \\
\hline
\hline
\end{tabular}
}
\par
\medskip
\begin{minipage}{0.97\linewidth}
Note --- The last column give a tentative classification; for the
  sources with more than one possible optical counterparts, this classification
  holds only for the actual counterpart.\\
$^a$ Probability of finding the number of observed sources or more sources inside the error circle, which is also the probability that all opticals object within the error circle are chance positional coincidences. \\
$^b$ X-ray luminosity derived in the 0.3-7 keV band assuming cluster-membership.  \\
$^c$ Absolute V-band magnitude assuming cluster-membership. The apparant V-band magnitude is computed from ($B_\mathrm{435}$ +  $r_{625}$)/2.\\
$^d$ Ratio of X-ray to optical ($r_{625}$) flux, using log
  ($f_\mathrm{X}$/$f_\mathrm{r}$) = log $f_\mathrm{X}$ + 5.67 + 0.4 $r_\mathrm{625}$ (Green et al. 2004); $f_\mathrm{X}$
  is derived in the 0.3-7 keV band.\\
$^e$ The classifications of 2b and 2c are valid only if the counterparts are true. CV: cataclysmic variable; AB: chromospherically active binary; 
chance counterpart: the optical source is located inside the error circle by chance
\end{minipage}
\end{sidewaystable*}

We first used the results of revised astrometry described above to search for probable \hst\ optical counterparts, 
which are within the 95\% confidence position error circles of the \chandra\ X-ray source positions. For the case of multiple optical 
sources inside the 95\% error circle, we included all of them as possible counterparts. 
Within the \hst\ ACS field of view, there are five X-ray sources and each of them has 1-3 optical counterparts. 
We then calculated the probabilities of chance positional coincidences for the five X-ray sources and their possible optical counterparts. 
We calculated the average number of sources within the error circles by computing the area ratios of the error circles to the field of view of the 5$\arcsec$ $\times$ 5$\arcsec$ finding charts and also the total number of the sources inside the field of view of the 5$\arcsec$ $\times$ 5$\arcsec$ finding charts. 
With the assumption of Poisson distribution, we obtained the probabilities of finding the number of observed sources or more sources inside the error circles and listed them (chance 1) in Table 4. Based on the probability, for CX1, CX3, and CX5, it is possible that all the optical sources inside the error circles are located there by chance and are not secure optical counterparts. For CX2, because of the low probability ($\sim$0.1$\%$) of finding three or more sources inside the error circle, one of the three sources probably is the real optical counterpart. CX4 is a marginal case and we cannot exclude the possibility that neither of the two sources within the error circle is the real counterpart. We further checked the number of false counterparts matches by shifting all the \chandra\ positions in four directions (north, south, east, and west) and 5$\arcsec$ away from the sources. We again calculated the average number of false matches and the probabilities of getting the number of observed sources or more sources inside the error circles with Poisson distribution (chance 2). We summarized the results in table 4. The results are similar to the previously calculated probabilities and confirm them.

Apart from checking the positional coincidences, we then checked whether these sources have special behavior in the optical CMDs. 
For the optical candidate counterparts of the X-ray sources, we expect they will 
have unusual optical properties (e.g. the color of candidate counterparts) corresponding to different kinds of X-ray sources. 
Through inspecting the positions on the CMDs of the candidate counterparts, we can know more about their nature in the optical band 
which will help us identify the X-ray sources with higher confidence. 
Finally, we compared the optical characteristics with the X-ray properties and the X-ray to optical 
flux ratio $f_\mathrm{X}$/$f_\mathrm{OPT}$ (See table 3) to classify possible counterparts.

CX1 has the highest X-ray photon count rate and is the second most luminous source in the whole field, and it is the one at 0.3$\arcmin$ from the center of M12 (in a field of view of 8.3$\arcmin$ $\times$ 8.3$\arcmin$). The probability of one of the two brightest sources of M12 occurring within the 0.3$\arcmin$ half-mass radius is only $\sim$0.8$\%$, and we  conclude that CX1 is a secure member of M12 - independent on the optical identification. This implies an X-ray luminosity (See table 3) which is typical for a CV, and rather too high for an AB of main-sequence stars. If it were an AB, Figure 6 shows that it would be the most X-ray luminous AB so far detected in any globular cluster and it would be more than 100 times brighter than any AB at the same optical magnitude near the Sun (100 if the optical star is the counterpart, more than 100 if the optical counterpart is fainter). We conclude that it is a CV. Its X-ray spectrum and its variability (Figure 3) on a time scale of an hour are compatible with this conclusion. The possible optical counterpart of CX1 shows no special color or H$\alpha$ emission in the \hst\ ACS observation. In the WFPC2 observation, no significant color change is detected. The optical counterpart shows no CV characteristics. Therefore, it is possible that the optical source inside the \chandra\ error circle may not be the real counterpart of CX1.

We included 3 optical counterpart candidates of CX2. CX2 has relatively hard X-ray color. 
For the optical counterpart candidates, CX2b have a relatively low X-ray to optical flux ratio and is redder and below the subgiant branch of the CMDs. 
The position of CX2b on the CMD is consistent with some identified sub-subgiants in other globular clusters (e.g. Mathieu et. al 2003; optical sources No.14 and No.43 in figure 9 of Edmonds et al. 2003a). 
The position of CX2b on figure 6 is also consistent with an AB. 
Therefore, we suggest that CX2b is an AB if it is the optical counterpart. 
In the case of CX2c, it is obvious bluer than the other two and has some H$\alpha$ emission, so it is a 
possible CV. It is located above the line that represent the maximum X-ray luminosities of ABs in globular clusters on figure 6, 
and its Xray to optical ratio indicates that it is more likely to be a CV rather than an AB. Hence, we classify CX2c as a CV.
As for CX2a, it is located on the main sequence of the CMD from the ACS observation and also has relatively low X-ray to optical flux ratio, while it is slightly redder that the main sequence during the WFPC2 observation.
From figure 6, it is consistent with an AB in globular clusters. Thus, CX2a might be a possible AB. 
However, without further information, we cannot make a secure classification.

CX3 also has a relatively hard spectral property.
The only one optical counterpart candidate of CX3 is an optically faint source, which makes its X-ray to optical ratio much higher than others. 
It is almost on the detection limit of the R-band observation. Besides, the blue color of CX3 indicates that it is more possible for CX3 to be an AGN instead of an AB (See figure 4). 
Furthermore, the relatively high X-ray to optical ratio and its position on the X-ray 
luminosity to absolute V-band magnitude diagram (Figure 6) show that CX3 could be a possible AGN if it is the counterpart. 
Nevertheless, the probability of chance coincidence of CX3 suggests that the optical source is very likely not a real counterpart. 
Therefore, we cannot have a confident classification for CX3.

In the case of CX4, the X-ray color indicates a relatively soft spectrum. CX4b is brighter and does not show any unusual emission 
on the CMDs. According to the X-ray to optical flux ratio, it could be a probable AB but it is a less confident identification. 
Turning to CX4a, it is also located on the main sequence of the CMDs. 
Similar to CX4b, CX4a could be a probable AB when considering the relatively soft X-ray colors and its X-ray to optical ratio (figure 6). 
Also because of the insufficient information, the classification of CX4a is less secure.

CX5 has two counterpart candidates inside its error circle, both lie on the main sequence. The X-ray luminosity of CX5 is higher than expected for active binaries at the absolute magnitude of CX5a, and also, albeit less so, for CX5b. If CX5 is a CV, we may expect to see a blue optical counterpart with H$\alpha$ emission, contrary to what is observed. Perhaps the two optical objects are not the counterpart. Possibly CX5 is a background AGN with a very faint optical counterpart. For the moment we cannot classify CX5.

\hst\ ACS and WFPC2 observations only cover above five sources of the observation. By comparing the position of the \chandra\ X-ray sources with 
previous optical observations (von Braun et al. 2002; Geffert et al. 1991), 
there is no other possible candidate counterpart for the rest of the X-ray sources.
Therefore we only describe the X-ray properties for the other sources not covered by ACS field of view, and also describe the optical properties for some of the X-ray sources with possible WFI optical counterparts.  
CX6 is inside the half-mass radius of M12 but outside the FOV of ACS. 
The WFI image shows a point-like candidate counterpart to CX6 near a saturated source. 
As a result, the possible contamination from the nearby source makes the shape of the CX6 candidate counterpart non-circular.
The X-ray color of CX6 is relatively soft compared with other sources. 
Furthermore, its luminosity (assuming it is a cluster member) is roughly 4 $\times$ $10^{30}$ \,ergs\,s$^{-1}$ , a very faint X-ray source. 
However, we can not classify CX6 without further information.
Regarding other sources outside the half-mass radius, the brightest one CX7 has similar X-ray colors with CX1 
which are relatively hard, so it could be a CV or an AGN. 
Remaining \chandra\ sources are faint and have luminosity lower than $10^{32}$ \,ergs\,s$^{-1}$. 
We found only CX12 coincided with a point-like source on the WFI image. 
However, we do not have enough information such as optical characteristics for identification. 

We summarized our identification in table 3. The classification of X-ray sources for CX2 holds only for the real optical counterpart.

\begin{inlinetable}
\caption{Probability of Chance Coincidence}
\begin{tabular}{lccc}
\hline 
\hline
CX & chance 1($\%$)\tablenotemark{a} & N\tablenotemark{b} & chance 2($\%$)\tablenotemark{c} \\
\hline
1 & 18 & 0.5 & 39\\
2 & 0.1 & 0.25 & 0.2\\
3 & 37 & 0.5 & 39\\
4 & 1 & 0.25 & 3\\
5 & 17 & 0.25 & 21\\
\hline
\hline
\end{tabular}
\par
\medskip
\begin{minipage}{0.82\linewidth}
NOTES. --- \\
$^a$ Probability of finding the number of observed sources or more sources inside the error circle, which is also the probability that all opticals object within the error circle are chance positional coincidences. \\
$^b$ N is the average number of false counterparts matches by shifting all the Chandra positions in four directions (north, south, east, and west) and 5$\arcsec$ away from the sources. \\
$^c$ Probabilities of getting the number of observed sources or more sources inside the Chandra error circles with Poisson distribution.
\end{minipage}
\end{inlinetable}

\begin{table*}
\centering{\small
    \caption{Scaling Parameters of Globular Clusters\label{tab:scaling}}
\begin{tabular}{lccccccc}
    \hline \hline
      Cluster &
      log $\rho_0$ & $r_\mathrm{c}$ & $d$ & $M_V$ & $\Gamma$ & $M_\mathrm{h}$ & $N_\mathrm{S}$\\ 
      &
      ($L_\sun\ \mathrm{pc}^{-3}$) &
      (\arcsec) &
      (kpc) &&&\\
\hline
47 Tuc	 & 4.81 &	24.0	& 4.5 &	-9.40 &	24.91 &	10.00 & \\
NGC6752 &	4.91 &	10.2 &	4.0 &	-7.73 &	5.02 &	2.15 & \\
M4 &	4.01 &	49.8 &	1.7 &	-6.90 &	1.00 &	1.00 & \\
M55 &	2.15 &	169.8 &	5.3 &	-7.60 &	0.18 &	1.82 & 3\\
NGC6366 &	2.42 &	109.8 &	3.6 &	-5.80 &	0.09 &	0.35 & 2\\
NGC288 &	1.80 &	85.2 &	8.8 &	-6.70 &	0.04 &	0.83 & 2\\
M12 &	3.23 &	43.2 &	4.9 &	-7.32 &	0.41 &	1.47 & 2\\
\hline
\hline
\end{tabular}
}
\par
\medskip
\begin{minipage}{0.85\linewidth}
Note --- Values for central density ($\rho_0$), core-radius
      ($r_\mathrm{c}$), distance ($d$) and absolute visual magnitude
      ($M_V$) originate from Harris\,1996 (version of February
      2003). For M4, the values of $\rho_0$ and $M_V$ are computed for
      the distance and reddening of Richer et al.\,(1997). The
      collision number is computed from $\Gamma \equiv \rho_0^{1.5}\
      r_\mathrm{c}^2$ and the half-mass from $M_\mathrm{h} \equiv
      10^{-0.4M_V}$. Values for $\Gamma$ and $M_\mathrm{h}$
      are normalized to the value of M4. $N_\mathrm{S}$ is the minumum number of secure X-ray sources in globular clusters (Bassa et al. 2008).
\end{minipage}
\end{table*}

\section{Discussion}
We identified 5 \chandra\ X-ray sources inside the \textit{HST} ACS field of view based on their optical counterparts. 
According to previous discussion ($\S2.2$), the expected background sources are 3-5 out of 6 sources inside the half-mass radius of M12. 
Thus, we cannot exclude the possibility (roughly 38$\%$) that all of the X-ray sources inside the half-mass radius are unrelated to the globular cluster. 
However, among these sources, if further considering the properties of their optical counterparts, 
CX1, CX2 are highly probable cluster members of M12 because of its unusual colors, while CX3, CX4 and CX5 are less secure members. 
The highly probable cluster members consist of one, possibly two (if 2c is the real counterpart) CVs and possibly one (if 2b is the real counterpart) AB.

From the study of Pooley et al. (2003), the numbers of X-ray sources with their 
luminosity $L_\mathrm{X(0.5-6.0 keV)}$ $\gtrsim$ 4 $\times$ $10^{30}$ \,ergs\,s$^{-1}$ have a strong correlation with the encounter rate $\Gamma$ of the 
globular cluster. The encounter rate is derived from $\Gamma$ $\equiv$ $\rho_{0}^{1.5}r_\mathrm{c}^{2}$ where $\rho_{0}$ is the central density and $r_\mathrm{c}$ is the core radius of the globular cluster (Verbunt 2003). $\Gamma$ describes the chance of close encounters or tidal captures between two objects in a globular cluster. 
As a result, $\Gamma$ provides the probability of dynamical formation of the X-ray sources in globular clusters.

On the other hand, we consider the probability of primordial formation channel that the number of X-ray sources scales with the mass of globular clusters.
Bassa et al. (2008) suggest an X-ray source number dependence on the encounter rate and the mass as N = 1.2$\Gamma_\mathrm{(M4)}$ + 1.1$M_\mathrm{h(M4)}$. This correlation is fitted with imposed minimum numbers of sources for some low core density globular clusters, NGC 288, M55, and NGC6366. The lower limit  to the number of secure members of M12 is also listed (See table 5). 
If we compute the expected value for M12 using the above relation, we obtain 0.5 and 1.6 due to encounters and primordial, respectively, in excellent agreement with the observed 2 secure members with $L_X$ above the limit used by Pooley et al. (2003). It is probable that both CX1 and CX2 are primordial binaries; alternatively, one may be a product of a close encounter. It is unlikely that both CX1 and CX2 are the product of a close encounter.

\begin{acknowledgements}
We thank Andrew Dolphin for providing a modified version of his \textsc{DOLPHOT} code.
This project is supported by the National Science Council of the Republic of China (Taiwan) through grant NSC 96-2112-M-007-037-MY3.
This research is based on observations made with the NASA/ESA Hubble
Space Telescope, obtained from the Data Archive at the Space Telescope
Science Institute, which is operated by the Association of
Universities for Research in Astronomy, Inc., under NASA contract NAS
5-26555. These observations are associated with program $\#$10005 and $\#$8118.
\end{acknowledgements}

\end{document}